# Superconductivity with $T_c$ 116K discovered in antimony polyhydrides

K. Lu[1,2], X. He[1,2,3] C.L. Zhang[1,2], Z.W. Li[1,2], S.J. Zhang[1], B. S. Min[1,2], J. Zhang[1,2], J.F. Zhao[1,2], L.C. Shi[1,2], Y. Peng[1,2], S.M. Feng[1], Q.Q. Liu[1], J. Song[1,2], R.C. Yu[1,2], X.C. Wang*[1,2], Y. Wang[4], M. Bykov[4], C. Q. Jin*[1,2,3]

[1] *Beijing National Laboratory for Condensed Matter Physics, Institute of Physics, Chinese Academy of Sciences, Beijing 100190, China*
[2] *School of Physical Sciences, University of Chinese Academy of Sciences, Beijing 100190, China*
[3] *Songshan Lake Materials Laboratory, Dongguan 523808, China*
[4] *Institute for Geowissenschaften, Johann Wolfgang Goethe University, Germany*

## ABSTRACT

Superconductivity (SC) was experimentally observed for the first time in antimony polyhydride. The diamond anvil cell combined with laser heating system was used to synthesize the antimony polyhydride sample at high pressure and high temperature conditions. *In-situ* high pressure transport measurements as function of temperature with applied magnet are performed to study the SC properties. It was found that the antimony polyhydride samples show superconducting transition with critical temperature $T_c$ 116 K at 184 GPa. The investigation of SC at magnetic field revealed that the superconducting coherent length ~40 Å based on Ginzburg Landau (GL) equation. Antimony polyhydride superconductor has the second highest $T_c$ in addition to sulfur hydride among the polyhydrides of elements from main group IIIA to VIIA in periodic table.

*Corresponding authors: wangxiancheng@iphy.ac.cn; jin@iphy.ac.cn

# INTRODUCTION

Recently interests are growing rapidly to explore new high temperature superconductors in polyhydrides based on the scenario that precompression effect will reduce the hydrogen metallization pressure to an experimentally accessible scope [1-11]. After the sulfur hydrides were theoretically predicted to host high temperature superconductivity (SC) [4, 5], it was soon experimentally discovered that $SH_3$ did exhibit SC with transition temperature $T_c$ about 203 K at 155 GPa [12]. This report of high $T_c$ superconductor of $SH_3$ accelerated the process of experimentally searching for other binary hydride superconductors [13-26], such as the discovery of SC at megabar pressure in $LaH_{10}$ with $T_c$ = 250-260 K[13, 15], $YH_9$ with $T_c$ = 243-262 K[17, 27] and $CaH_6$ with $T_c$ of ~210 K[18, 19]. Besides the rare earth polyhydride and alkali earth polyhydride superconductors with $T_c$ exceeding 200 K, other metal polyhydrides with moderate $T_c$ have also been experimentally reported [16, 20-26], such as $ThH_{10}$ with $T_c$ = 161 K at 175 GPa [20]. For the heavy rare earth elements with $f$ shell fully filled, the lutetium hydride of $Lu_4H_{23}$ was reported to exhibit SC with $T_c$ = 71 K at 218 GPa [21]. The hydrides of IVB and VB group metal of Zr, Hf and Ta were found to host SC at megabar pressure with $T_c$ = 71 K [22], 83 K [23] and 30 K [26], respectively. Most of discovered binary hydride superconductors are found to be located in group IIA and IIIB in the periodic table and have the electronegative values smaller than 1.5. It is a fundamental consideration that the metal element with small electronegativity can provide electrons to occupy the anti-bonding orbital of hydrogen in order to prevent the formation of hydrogen molecular during the hydrogen metallization process upon the compression. However for the elements from main group IIIA to VIIA in the periodic table, only a few covalently bonded hydride superconductors have been experimentally discovered. Beside SH$_3$ with $T_c$ =203 K, SnH$_n$ was reported to have SC with $T_c$ about 70 K [25], and phosphorus hydride was found to be superconducting with $T_c$ about 103 K by using PH$_3$ as the precursor [24, 28]. Here we report an experimental discovery of another covalently bonded hydride superconductor of antimony polyhydride superconductor. The SC with $T_c$ = 116 K was experimentally observed that is the second highest $T_c$ so far reported in the polyhydrides of elements in main group IIIA to VIIA in the periodic table.

## EXPERIMENTAL DETAILS

The antimony polyhydrides were synthesized at high pressure and high temperature conditions based on the diamond anvil cell high pressure techniques in combination with laser heating technique. The diamond anvils with the culet diameter of 50 μm beveled to 300 μm were used for the megarbar pressure experiments. The gasket made of T301 stainless was prepressed to ~10 μm in thickness, and drilled with a hole of 300 μm in diameter. Then the hole was filled with aluminum oxide that was densely pressed before further drilled to a hole of 40 μm in diameter serving as sample chamber. The ammonia borane (AB) was filled into the high pressure chamber to act both as the hydrogen source as well as the pressure transmitting medium. The Pt was deposited on the surface of the anvil culet to serve as the inner electrodes. An antimony foil with the size of 20 μm(L) ∗ 20 μm(W) ∗ 1 μm(T) was stacked on the inner electrodes. The pressure was calibrated by the shift of Raman peak of diamond. The details are described in the ATHENA procedure reported in Ref.[36].

A YAG laser with a wavelength of 1064 nm was used to *In-situ* heat the high pressure sample. The laser beam size is about 5 μm in diameter. The sample was laser heated at 2000 K for several minutes during which hydrogen released from the AB would react with the antimony to form antimony polyhydride. The high temperature was determined by fitting the black body irradiation spectra. The synthesis pressure was kept unchanged after the synthesis further for electric transport measurements. The high pressure electric conductivity experiments were performed in a MagLab system with temperatures from 300 K to 1.5 K and a magnetic field up to 5 Tesla. A Van der Pauw method was employed as the general high pressure resistance measurements[37, 38] while the applied electric current is set to be 1 mA.

The detail of *in-situ* high pressure x-ray diffraction experiments can be seen in the online supplementary material.

## RESULTS AND DISCUSSIONS

The sample was synthesized at 184 GPa with the sample chamber & electrodes assembly shown in Fig.S 1. The temperature dependence of resistance measured at the same pressure is shown in Fig. 1. The resistance decreases smoothly with temperature decrease before it drops sharply at the onset temperature 116 K and reaches zero gradually. The zero resistance is presented clearly in the inset of Fig. 1, which rules out the possibility of the resistance drop with an origin from the structural or magnetic phase transitions. Therefore it is suggested that a superconducting transition happens. To clearly determine the transition temperature, the resistance derivative over temperature is plotted in the inset of Fig. 1. The derivative curve shows a sharp peak, and onset superconducting transition temperature $T_c$ can be determined to be 116 K by the right upturn temperature. The kinks during the resistance dropping suggest multistep superconducting transitions, which are possibly caused by the generated antimony polyhydrides with different hydrogen content as can be generally seen in other polyhydride superconductors [13, 18, 19].

The dependence of the transition on magnetic field was studied. As shown in Fig. 2, the transition temperature is suppressed gradually by applying magnetic field, which is consistent with the superconducting properties and further confirms the nature of superconducting transition. The $T_c^{90\%}$ values at different magnetic fields were determined by the criteria of temperature where the resistance drops to 90% relative to the normal state at the onset temperature as shown by the dashed line in Fig. 2. The synthesis of polyhydride antimony is rather challenging because it is extremely difficult to heat the sample by laser. The synthesis can only be carried out with high laser power that frequently damages the anvils. This prevents us from further increasing or releasing pressure to study the pressure dependence of SC as we have performed for other polyhydride superconductor studies [18, 26]. The zero resistance at zero field is suppressed by applying magnetic field. It is speculated that there should exist weak superconducting links between the generated superconducting crystalline grains. Upon high magnetic field is applied and penetrates the superconducting sample, the weak superconducting links would be broken and suppress the zero resistance as observed in the granular superconductors [29-31].

The upper critical magnetic field $\mu_0 H_{c2}(T)$ versus temperature was plotted in Fig. 3, which presents a straight line. After linearly fitting the data, the slope of $|dH_c/dT|$ was obtained to be 0.25 T/K. This slope is significantly smaller than those of clathrate hydride superconductors, such as 1.73 T/K for $CaH_6$ ($T_c$~210 K) [18], 1.03 T/K for $LaH_{10}$ ($T_c$~250 K) [13], and 1.06 T/K for $Lu_4H_{23}$ ($T_c$~71 K) [21], while it is comparable with those of covalent bonding dominant hydride superconductors of $SH_3$ (~0.5 T/K, $T_c$~203 K) [32] and SnH$_n$ (0.21 T/K, $T_c$~71 K) [25]. It seems that the magnetic vortex pinning force in covalently bonded hydride superconductors is generally weaker than that for ionic bonding dominant clathrate types. According to the Werthamer-Helfand-Hohenberg (WHH) theory, the $\mu_0H_{c2}(0)$ controlled by orbital depairing mechanism in a dirty limit ($\mu_0H_{c2}^{Orb}(0)$) can be estimated with a formula of $\mu_0H_{c2}(T) = -0.69 \times [dH_{c2}/dT|_{Tc}] \times T_c$. Taking the slope of -0.25 T/K and $T_c^{90\%} = 115$ K, the $\mu_0H_{c2}^{Orb}(0)$ can be calculated to be ~20 T. The $\mu_0H_{c2}(0)$ can also be estimated by using the Ginzburg Landau (GL) theory with an equation of $\mu_0H_{c2}(T) = \mu_0H_{c2}^{GL}(0)(1-(T/T_c)^2)$. The equation was fitted using the $\mu_0H_{c2}(T)$ as shown in Fig. 3. The fitting yields the parameter of $\mu_0H_{c2}^{GL}(0)$ ~16 T that is comparable with $\mu_0H_{c2}^{Orb}(0)$. In addition the $\mu_0H_{c2}(0)$ limited by spin depairing mechanism associated with Zeeman effect ($\mu_0H_{c2}^{P}(0)$) for the case of weak coupling superconducting system is determined by the formula of $\mu_0H_{c2}^{P}(0) = 1.86 \times T_c$. The $\mu_0H_{c2}^{P}(0)$ can be calculated to be 211 T by using the $T_c^{90\%} = 115$ K. The small $\mu_0H_{c2}^{Orb}(0)$ value relative to $\mu_0H_{c2}^{P}(0)$ is indicative that the Cooper pair is broken through the orbital depairing mechanism. Finally, the GL coherent length $\xi$ is estimated to be ~40 Å by the equation of $\mu_0H_{c2}^{GL}(0)= \Phi_0/2\pi\xi^2$ where $\Phi_0= 2.067 \times 10^{-15}$ Web is the magnetic flux quantum.

The antimony hydride SC has been theoretically investigated [33-35]. In these studies only SbH and SbH$_4$ are predicted to be stable above 150 GPa, and SbH$_4$ was proposed to be SC with very high $T_c$ value of ~100 K at 150 GPa [34, 35]. SbH$_3$ however was only expected to be stable above 300 GPa with a very low superconducting $T_c$ ~20 K [35]. We speculate that our observed SC with $T_c$ ~116 K is probably from the $P6_3/mmc$-SbH$_4$ phase. We preliminarily investigated the superconducting phase by *in-situ* high pressure *x*-ray experiments. The sample was synthesized at 208 GPa using the standard symmetric diamond anvil cell specifically for synchrotron radiation measurements. The diffraction pattern presented in Fig. S2 shows the possible existence of the hexagonal phase of $SbH_4$.

For the binary hydrides with the elements located in main group IIIA to VIIA of the period table, covalent bonding is usually dominant between the element and hydrogen since their electronegativity values are comparable. The typical example is $SH_3$, where strong polar covalent bonding was proposed between the adjacent S and H atoms by the calculated electron localization function (ELF) [5]. For $SbH_4$ the calculated ELF at 150 GPa also presents a covalent bonding between neighbor Sb and H atoms [34]. Hence $SbH_4$ is another experimentally reported covalently bonded high $T_c$ hydride superconductor in addition to $SH_3$. According to the structure model of $SbH_4$ at 150 GPa [34], the schematic view of the crystal structure is plotted as shown in Fig. S3a. There are two Wyckoff positions for hydrogen atoms: H1 (4e) and H2 (4f) denoted with yellow and green colors, respectively. H1 atoms are located in the octahedral interstitial sites of Sb lattice (Fig. S3b) while H2 atoms in the tetrahedral interstice (Fig. S3c). Unlike $SH_3$, H1 atoms in $SbH_4$ form quasi hydrogen molecules with the bonding length of 0.83 Å (Fig. S3a), while the second shortest H-H distance is ~1.73 Å. In $SH_3$ the H-H covalent bond is considered to be absent due to the large H-H distance (1.49 Å), while the metalized S-H valence bond is believed to be responsible for the high $T_c$ SC. The larger second shortest H-H distance in $SbH_4$ relative to the shortest one in $SH_3$ implies that $SbH_4$ should have a weaker H-H bonding strength and thus lead to a relatively weaker electron-phonon coupling. That is why $SbH_4$ has a lower $T_c$ than $SH_3$ has. Despite all this antimony binary hydride experimentally shows a high temperature SC with $T_c$ exceeding 110 K, the second highest so far for the polyhydride compounds of elements from main group IIIA to VIIA in the periodic table.

## CONCLUSION

In summary the antimony polyhydride superconductor has been experimentally discovered. The antimony polyhydride shows SC with $T_c \sim 116$ K at 184 GPa. The upper magnetic field is $\mu_0 H_{c2}(0) \sim 20$ T with a GL coherent length ~40 Å.

**Figure Captions:**

Fig. 1 The temperature dependence of resistance for antimony polyhydride sample measured at 184 GPa. The upper inset is the derivative of resistance over temperature to clearly show the superconducting onset temperature ($T_c$). The lower inset demonstrates the zero resistance at low temperature.

Fig. 2 The temperature dependence of resistance measured at different magnetic fields.

Fig. 3 The upper critical magnetic field $\mu_0 H_{c2}(T)$. The red line represents the GL fitting. The inset displays the linear fitting.

**Fig 1**

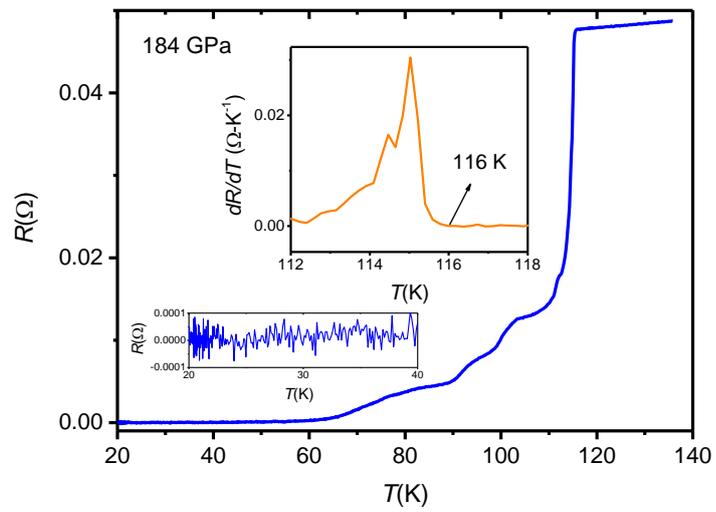

**Fig. 2**

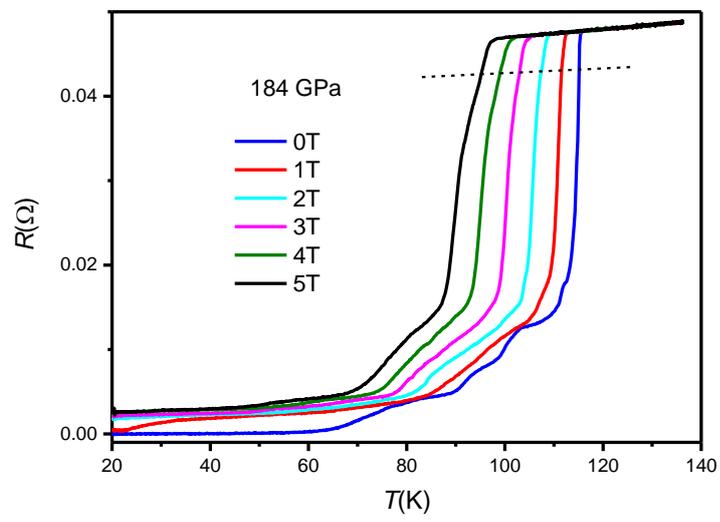

**Fig. 3**

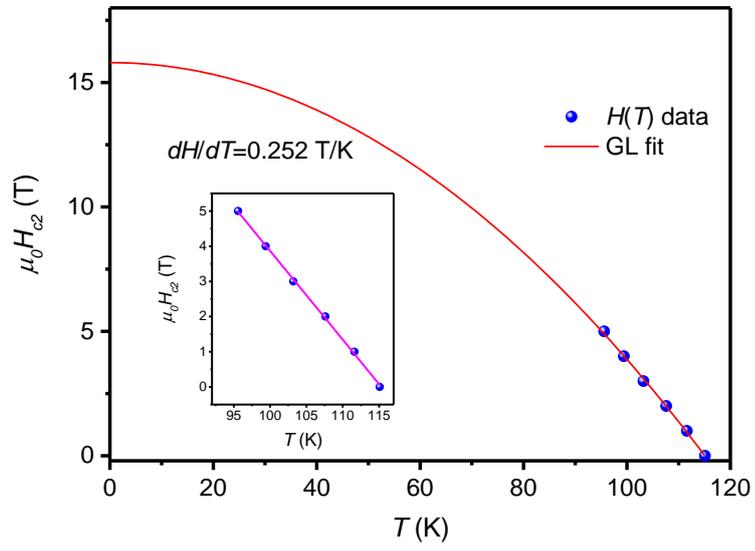